\documentclass[%
    final,
    notitlepage,
    narroweqnarray,
    inline,
    twoside,
    ]{ieee}

\usepackage{graphicx}
\usepackage{array}

\begin{document}

\title[Measurements of integral muon intensity at large zenith angles]{%
       Measurements of integral muon intensity \\at large zenith angles}

\author{A.N.~Dmitrieva, D.V.~Chernov, R.P.~Kokoulin, K.G.~Kompaniets,
G.~Mannocchi, A.A.~Petrukhin,\\ O.~Saavedra, V.V.~Shutenko,
D.A.~Timashkov, G.~Trinchero, I.I.~Yashin
        \authorinfo{A.N.~Dmitrieva,
D.V.~Chernov, R.P.~Kokoulin, K.G.~Kompaniets, A.A.~Petrukhin,
V.V.~Shutenko, D.A.~Timashkov and I.I.~Yashin are with Moscow
Engineering Physics Institute (MEPhI), Moscow 115409, Russia.
{E-mail}: ANDmitriyeva@mephi.ru.}%
\and{} \authorinfo{G.~Mannocchi and G.~Trinchero are with Istituto
Nazionale di Astrofisica, Sezione di Torino, 10133 Torino, Italy.}
\and{} \authorinfo{O.~Saavedra is with Dipartimento di Fisica
Generale dell Universita di Torino, 10125 Torino, Italy.} }

\firstpage{1}

\maketitle

\begin{abstract}
High-statistics data on near-horizontal muons collected with
Russian-Italian coordinate detector DECOR are analyzed. Precise
measurements of muon angular distributions in zenith angle
interval from 60$^\circ$ to 90$^\circ$ have been performed. In
total, more than 20~million muons are selected. Dependences of the
absolute integral muon intensity on zenith angle for several
threshold energies ranging from 1.7~GeV to 7.2~GeV are derived.
Results for this region of zenith angles and threshold energies
have been obtained for the first time. The dependence of integral
intensity on zenith angle and threshold energy is well fitted by a
simple analytical formula. \\ {}
\end{abstract}


\section{Introduction}

\PARstart Studies of angular and energy dependence of muon flux at
the Earth's surface give important information as about processes
of muon generation and propagation in the atmosphere so about
primary cosmic rays. Measurements of muon flux at large zenith
angles up to 90$^\circ$ are especially actual since primary
particles for such muons have higher energies than in the vertical
direction. Experimental studies of muon intensity at large zenith
angles at the ground level can be conditionally separated in two
groups: measurements of muon integral intensity with threshold
energies less than 1~GeV [1]--[8] and investigations of integral
and differential muon spectra for muon energies higher than 10~GeV
(see review [9]). Regions of measurements of muon spectrum at
large zenith angles are presented in Fig.~1. It is remarkable that
for threshold energies from 1~GeV to 10~GeV and zenith angles
60$^\circ{\le}{\theta}{\le}90^\circ$ muon intensity data are
absent.

\begin{figure}[h]
\centerline{\includegraphics[scale=0.8]{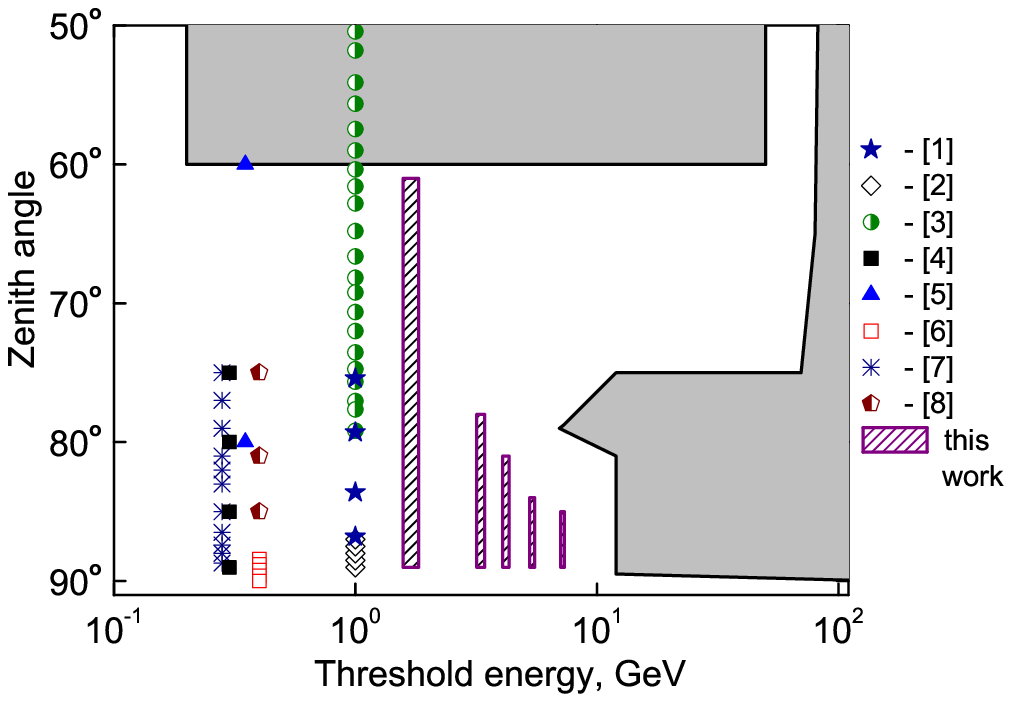}}
\caption{Regions of muon spectrum measurements at large zenith
angles. Symbols represent the measurements of integral intensity;
shaded areas are the regions of differential spectrum
measurements. Dashed areas are the regions investigated in this
work.}
\label{fig01} 
\end{figure}

To explore this region, a setup capable to measure near-horizontal
muon flux at different threshold energies with a good angular
accuracy of track reconstruction is needed. Coordinate detector
DECOR, which is a part of experimental complex NEVOD situated in
MEPhI (Moscow), is such a detector. Regions of threshold energies
and zenith angles accessible for DECOR and analyzed in this work
are shown by the dashed areas in Fig.~1.

\section{Experimental setup and events selection}

Experimental complex NEVOD includes a water Cherenkov calorimeter
NEVOD [10] with sensitive volume 2000~m$^3$ equipped with
quasispherical modules of PMTs, and large-area ($\sim$~110~m$^2$)
coordinate detector DECOR [11] (Fig.~2). Eight supermodules (SM)
of DECOR are situated in the gallery around the water tank, and
four SM on its cover. SM of side part of DECOR represents eight
parallel planes with sensitive area $3.1$~m$~{\times}~2.7$~m,
suspended vertically with 6~cm distance from each other. These
planes consist of 16 chambers which contain 16 tubes with inner
cross-section $0.9$~cm$~{\times}~0.9$~cm. Chambers are operated in
a limited streamer mode and are equipped with two-coordinate
external strip read-out system. Thus, coordinates of passing
particle can be obtained for each plane with spatial accuracy of
muon track location $\sim$~1~cm. First level trigger is formed
when there are at least two even and two odd triggered planes in a
given SM.

\begin{figure}[h]
\centerline{\includegraphics[scale=0.6]{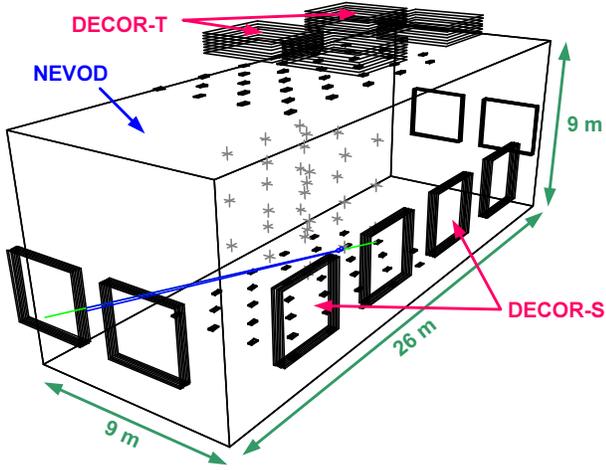}}
\caption{Experimental complex NEVOD-DECOR.}
\label{fig02} 
\end{figure}

For the analysis, particles passing through two SM situated at
different sides of the water pool were selected. Different pairs
of SM correspond to different values of threshold energy. Accuracy
of zenith angle reconstruction for tracks passing through selected
SM pairs is $0.3^{\circ}-0.5^\circ$. Selection procedure includes
the following conditions.

\begin{itemize}
\item "OneTrack" criterion: two tracks reconstructed from data of
different supermodules must coincide within 5$^\circ$ cone. In
this case the tracks in separate SMs are considered as tracks of
the same particle. Straight line connecting the middles of two
reconstructed track segments is taken as the trajectory of the
particle.

\item The events in which muon passed closer than 3~cm from the
boundary of SM are rejected in order to decrease the edge effects.

\item There must be two and only two track projections (X,Y) in
each SM for unambiguous reconstruction of geometrical
characteristics of muon track (the absence of accompanying
particles).
\end{itemize}

Data collected over a period from December 2002 to June 2003 are
analyzed. Total time of registration is equal to 3390~hours. The
total number of selected events is more than 20~millions.

\begin{table*}[ht]

\renewcommand{\arraystretch}{1.15}

\caption{Dependence of integral muon intensity on zenith angle}

\begin{center}
{\tt
\begin{tabular}{cccccc}\hline\hline

$\theta,^\circ$& \multicolumn{5}{c}{$I(\theta)\cdot10^5$,
(cm$^2{\cdot}$s${\cdot}$sr)$^{-1}$}\\ 

 &  1.7~GeV & 3.3~GeV &  4.2~GeV & 5.4~GeV & 7.2~GeV\\\hline

61&  $140.1\pm2.3$ &     &    &     &    \\

63&  $122.5\pm1.5$ &     &    &     &    \\

65&  $108.9\pm1.2$ &     &    &     &    \\

67& $93.6\pm0.9$ &     &    &     &    \\

69& $79.9\pm0.7$ &     &    &     &    \\

71&  $66.41\pm0.6$ &     &    &     &    \\

73&  $54.8\pm0.4$ &     &    &     &    \\

75& $44.0\pm0.3$ &     &    &     &    \\

77& $34.0\pm0.2$ &     &    &     &    \\

78& $29.6\pm0.2$ & $25.5\pm0.4$ &    &     &    \\

79&  $25.4\pm0.2$& $22.8\pm0.2$ &    &     &    \\

80& $21.5\pm0.2$ &$19.5\pm0.1$  &    &     &    \\

81& $18.2\pm0.1$ &$16.5\pm0.1$ &$15.7\pm0.4$&     &    \\

82& $14.9\pm0.1$ &$13.55\pm0.07$ &$12.9\pm0.2$&     &    \\

83& $11.98\pm0.09$ &$10.98\pm0.06$& $10.78\pm0.09$&     &
\\

84& $9.53\pm0.07$& $8.79\pm0.04$& $8.61\pm0.06$& $8.1\pm0.1$&
\\

85& $7.27\pm0.06$ &$6.74\pm0.03$ &$6.56\pm0.04$ &$6.55\pm0.08$
&$5.84\pm0.09$\\

86& $5.44\pm0.12$ &$5.08\pm0.02$ &$4.97\pm0.03$& $4.86\pm0.05$&
$4.59\pm0.04$\\

87& $3.95\pm0.09$ &$3.74\pm0.02$ &$3.65\pm0.02$ &$3.64\pm0.03$
&$3.51\pm0.02$\\

88& $2.78\pm0.06$& $2.64\pm0.01$ &$2.61\pm0.02$ &$2.56\pm0.02$
&$2.50\pm0.01$\\

89& $2.01\pm0.05$ &$1.83\pm0.01$ &$1.74\pm0.01$ &$1.77\pm0.01$
&$1.69\pm0.01$\\\hline\hline

\end{tabular}
}
\end{center}
\end{table*}

\section{Results}

Threshold energy $E_{\rm min}$ of muons passing through selected
pair of SM is calculated by means of range-energy tables [12]. It
is calculated for each selected event, and then the event is
placed in data array $N(\theta ,\varphi ,E_{\rm min})$. The bin of
zenith angle $\Delta{\theta}=1^\circ$, the bin of azimuth angle
$\Delta{\varphi}=0.5^\circ$,  the bin of threshold energy $E_{\rm
min}=250$~MeV. Integral muon intensity is calculated in the
following way:
\begin{eqnarray}
I(\theta,\varphi,E_{\rm min}) = {{N(\theta,\varphi ,E_{\rm min})}
\over {T \cdot\varepsilon^2_{\rm SM}\cdot \varepsilon _{\rm
MCS}\cdot \varepsilon _{\rm add} \cdot S\Omega (\theta ,\varphi
,E_{\rm min} )  }}, %
\label{one}
\end{eqnarray}
where $N(\theta ,\varphi ,E_{\rm min})$ is the number of
registered muons in a given angular and threshold energy bin. $T$
is "live time" of registration. The parameter $\varepsilon _{\rm
SM}$ is efficiency of single SM triggering, and $\varepsilon _{\rm
add}$ takes into account event rejection because of accompanying
particles. Results of simulations and additional experimental data
analysis give the following values: $\varepsilon _{\rm
SM}=0.936{\pm}0.004$, $\varepsilon _{\rm add}$ varies from 0.83 to
0.91 for different $\theta$ and $E_{\rm min}$ (uncertainty of
$\varepsilon _{\rm add}$ is less then 0.35\%). The function
$S\Omega (\theta ,\varphi ,E_{\rm min})$ is the setup acceptance
calculated by means  of MC method taking into account the
structure of SM and selection requirements.

\begin{figure}[h]
\centerline{\includegraphics[scale=0.9]{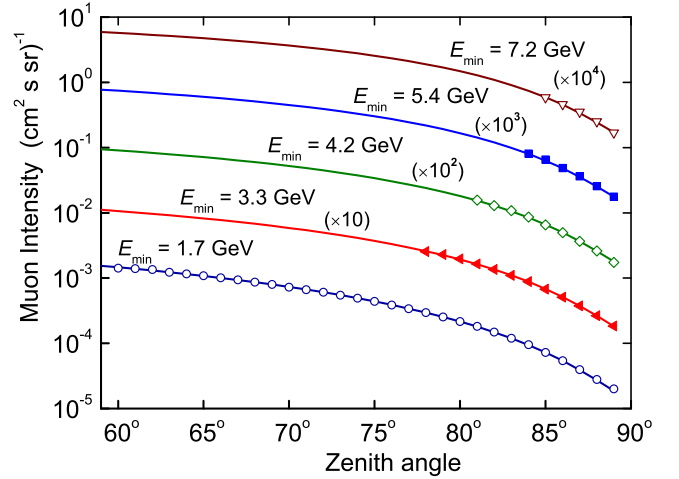}}
\caption{Dependence of absolute muon intensity on zenith angle at
several threshold energies. Symbols represent experimental results
obtained in this work. Curves are calculation results by formula
(2).}
\label{fig03} 
\end{figure}

Absolute muon intensity averaged in azimuth angle for zenith
angles 61$^\circ{\le}{\theta}{\le}89^\circ$ and for five threshold
energies is represented in Table~I and is shown in
Fig.~\ref{fig03} (points). Errors in the table include statistical
and systematical uncertainties (uncertainty of threshold energy
estimation, uncertainty of $\varepsilon _{\rm SM}$, muon energy
loss in the walls of surrounding buildings).

\section{Approximation formula}

For approximation of measured experimental data, the following
simple formula is used:
\begin{eqnarray}
{I_{\rm app}(\theta ,E_{\rm min}) = {{C_{\rm I}} \over
{E_1^\gamma}} \cdot~~~~~~~~~~~~~~~~~~~~~~~~~~~~~~~~~~~~{}
\nonumber } \\ {\cdot \exp \left(-{{\gamma}\over{\gamma+1}} \cdot
{{E_{\rm cr}}\over{E_1+a{\cdot}h_1}}  \cdot  \ln \left( {h_0 \over
{h_1{\cdot}\cos{\theta}^*}} \right) \right).}
\label{two}
\end{eqnarray}
The factor in front of the exponent reflects the form of muon
spectrum in the upper atmosphere, and the exponential function
takes into account muon decay. Here $C_{\rm I}$ is the
normalization; $E_1=E_{\rm min}+a\cdot(h_0/\cos \theta^*-h_1)$ is
the threshold muon energy (GeV) at production level. In this
formula $a=2.5{\cdot}10^{-3}$~GeV$\cdot$cm$^2$/g is effective
specific energy loss; $(h_0/\cos \theta^*-h_1)$ is the path of
muon in the atmosphere; $h_0=1018$~g/cm$^2$ is the total thickness
of atmosphere (altitude of setup under see level is taken into
account); $h_1=100$~g/cm$^2$ is the effective depth of muon
generation. $E_{\rm
cr}=z_0{\cdot}m/(c{\cdot}\tau_0{\cdot}\cos\theta^*)$ is the
effective critical energy for muon; $z_0$ is the effective length
at which the density of atmosphere is changed by a factor of $e$
[13]; $c$ is the velocity of light; $\tau_0$ is muon life time and
$m$ is it's mass (GeV);
$\cos\theta^*=({\cos}^{\alpha}{\theta}+{\Delta}^{\alpha})^{1/{\alpha}}$
is the approximation of effect of atmosphere sphericity. As a
result of fitting, the following values of free parameters were
obtained: $C_{\rm I}=0.103$, ${\gamma}=1.963$, $z_0=6.81$~km,
${\Delta}=0.0577$ and ${\alpha}=1.35$.

\begin{figure}[h]
\centerline{\includegraphics[scale=0.78]{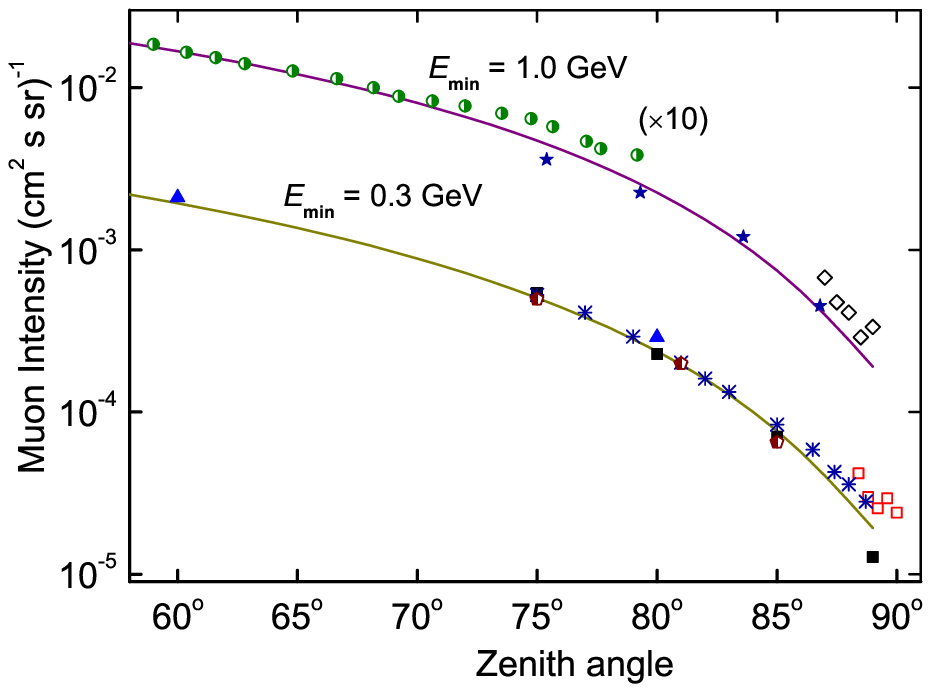}}
\caption{Dependence of absolute muon intensity on zenith angle for
lower thresholds (0.3 and 1~GeV). Curves are calculation results
by formula (2). Symbols for experiments are the same as in
Fig.~1.}
\label{fig04} 
\end{figure}

In Fig.~3, integral muon intensity calculated by formula (2) for
five threshold energies (the curves) is compared with the present
experimental data. Dependence of integral muon intensity on zenith
angle calculated for lower thresholds (1 and 0.3~GeV) and
experimental data of earlier measurements [1]--[8] are presented
in Fig.~4. Comparison of calculated values with data [1,4,5,7,8]
shows a reasonable agreement. In works [2] and [6], the intensity
is somewhat higher than measured in [1] or calculated by (2). The
integral intensity data at $E_{\rm min}=1$~GeV obtained in [3]
decrease with the increase of zenith angle more slowly than it
follows from [1] and calculation by (2), but at angles less than
72$^\circ$ the agreement is quite well.

\section{Conclusions}

Experimental data of coordinate detector DECOR cover unexplored
earlier region for integral muon intensity at threshold energies
$1.7{\le}E_{\rm min}{\le}7.2$~GeV and zenith angles
61$^\circ{\le}{\theta}{\le}89^\circ$. It is important to mark that
the measurements for all thresholds were performed simultaneously
with a single setup, that minimizes systematic uncertainties.
Extrapolation of the present data to lower thresholds is in a
reasonable agreement with the results of other measurements.

\section*{Acknowledgments}
The research is performed at the Experimental Complex NEVOD with
the support of the Federal Agency of Education and Federal Agency
for Science and Innovations (contracts 02.452.11.7064,
02.434.11.7039; program of support of leading scientific schools
10113.2006.2).

\end{document}